\documentclass[aps,prc,twocolumn,showpacs,superscriptaddress]{revtex4-1}

\usepackage{amsmath,amssymb,amsfonts}
\usepackage{graphicx,subfigure,color}

\draft 

\begin{document}

\preprint{AIP}

\title{Study of Three-Nucleon Dynamics in the dp breakup  collisions  using the WASA detector} 



\newcommand*{\IKPUU}{Division of Nuclear Physics, Department of Physics and 
 Astronomy, Uppsala University, Box 516, 75120 Uppsala, Sweden}
\newcommand*{\ASWarsN}{Nuclear Physics Division, National Centre for 
 Nuclear Research, ul.\ Hoza~69, 00-681, Warsaw, Poland}
\newcommand*{\IPJ}{Institute of Physics, Jagiellonian University, prof.\ 
 Stanis{\l}awa {\L}ojasiewicza~11, 30-348 Krak\'{o}w, Poland}
\newcommand*{\Edinb}{School of Physics and Astronomy, The University of
  Edinburgh, James Clerk Maxwell Building, Peter Guthrie Tait Road, Edinburgh
  EH9 3FD, Great Britain}
\newcommand*{\MS}{Institut f\"ur Kernphysik, Westf\"alische 
 Wilhelms--Universit\"at M\"unster, Wilhelm--Klemm--Str.~9, 48149 M\"unster, 
 Germany}
\newcommand*{\ASWarsH}{High Energy Physics Division, National Centre for 
 Nuclear Research, ul.\ Hoza~69, 00-681, Warsaw, Poland}
\newcommand*{\Budker}{Budker Institute of Nuclear Physics of SB RAS, 
 11~Acad.\ Lavrentieva Pr., Novosibirsk, 630090 Russia}
\newcommand*{\Novosib}{Novosibirsk State University, 2~Pirogova Str., 
 Novosibirsk, 630090 Russia}
\newcommand*{\PGI}{Peter Gr\"unberg Institut, PGI--6 Elektronische 
 Eigenschaften, Forschungszentrum J\"ulich, 52425 J\"ulich, Germany}
\newcommand*{\DUS}{Institut f\"ur Laser-- und Plasmaphysik, Heinrich Heine 
 Universit\"at D\"usseldorf, Universit\"atsstr.~1, 40225 Düsseldorf, Germany}
\newcommand*{\IFJ}{The Henryk Niewodnicza{\'n}ski Institute of Nuclear 
 Physics, Polish Academy of Sciences, ul.\ Radzikowskiego~152, 31-342 
 Krak\'{o}w, Poland}
\newcommand*{\PITue}{Physikalisches Institut, Eberhard Karls Universit\"at 
 T\"ubingen, Auf der Morgenstelle~14, 72076 T\"ubingen, Germany}
\newcommand*{\Kepler}{Kepler Center for Astro and Particle Physics,
  Physikalisches Institut der Universit\"at T\"ubingen, Auf der 
 Morgenstelle~14, 72076 T\"ubingen, Germany}
\newcommand*{\IKPJ}{Institut f\"ur Kernphysik, Forschungszentrum J\"ulich, 
 52425 J\"ulich, Germany}
\newcommand*{\ZELJ}{Zentralinstitut f\"ur Engineering, Elektronik und 
 Analytik, Forschungszentrum J\"ulich, 52425 J\"ulich, Germany}
\newcommand*{\Erl}{Physikalisches Institut, Friedrich--Alexander Universit\"at
 Erlangen--N\"urnberg, Erwin--Rommel-Str.~1, 91058 Erlangen, Germany}
\newcommand*{\ITEP}{Institute for Theoretical and Experimental Physics named 
 by A.I.\ Alikhanov of National Research Centre ``Kurchatov Institute'', 
 25~Bolshaya Cheremushkinskaya Str., Moscow, 117218 Russia}
\newcommand*{\Giess}{II.\ Physikalisches Institut, 
 Justus--Liebig--Universit\"at Gie{\ss}en, Heinrich--Buff--Ring~16, 35392 
 Giessen, Germany}
\newcommand*{\IITI}{Discipline of Physics, Indian Institute of Technology 
 Indore, Khandwa Road, Indore, Madhya Pradesh 453 552, India}
\newcommand*{\HepGat}{High Energy Physics Division, Petersburg Nuclear Physics 
 Institute named by B.P.\ Konstantinov of National Research Centre ``Kurchatov 
 Institute'', 1~mkr.\ Orlova roshcha, Leningradskaya Oblast, Gatchina, 188300
 Russia}
\newcommand*{\HeJINR}{Veksler and Baldin Laboratory of High Energiy Physics, 
 Joint Institute for Nuclear Physics, 6~Joliot--Curie, Dubna, 141980 Russia}
\newcommand*{\Katow}{August Che{\l}kowski Institute of Physics, University of 
  Silesia, ul.\ 75 Pu{\l}ku Piechoty 1, 41-500 Chorz\'{o}w, Poland}
\newcommand*{\NITJ}{Department of Physics, Malaviya National Institute of 
 Technology Jaipur, JLN Marg, Jaipur, Rajasthan 302 017, India}
\newcommand*{\JARA}{JARA--FAME, J\"ulich Aachen Research Alliance, 
 Forschungszentrum J\"ulich, 52425 J\"ulich, and RWTH Aachen, 52056 Aachen, 
 Germany}
\newcommand*{\Bochum}{Institut f\"ur Experimentalphysik I, Ruhr--Universit\"at 
 Bochum, Universit\"atsstr.~150, 44780 Bochum, Germany}
\newcommand*{\IITB}{Department of Physics, Indian Institute of Technology 
 Bombay, Powai, Mumbai, Maharashtra 400 076, India}
\newcommand*{\Tomsk}{Department of Physics, Tomsk State University, 36~Lenin 
 Ave., Tomsk, 634050 Russia}
\newcommand*{\KEK}{High Energy Accelerator Research Organisation KEK, Tsukuba, 
 Ibaraki 305--0801, Japan} 
\newcommand*{\ASLodz}{Astrophysics Division, National Centre for Nuclear
 Research, Box~447, 90-950 {\L}\'{o}d\'{z}, Poland}

\author{P.~Adlarson}    \affiliation{\IKPUU}
\author{W.~Augustyniak} \affiliation{\ASWarsN}
\author{W.~Bardan}      \affiliation{\IPJ}
\author{M.~Bashkanov}   \affiliation{\Edinb}
\author{F.S.~Bergmann}  \affiliation{\MS}
\author{M.~Ber{\l}owski}\affiliation{\ASWarsH}
\author{A.~Bondar}      \affiliation{\Budker}\affiliation{\Novosib}
\author{M.~B\"uscher}   \affiliation{\PGI}\affiliation{\DUS}
\author{H.~Cal\'{e}n}   \affiliation{\IKPUU}
\author{I.~Ciepa{\l}}   \affiliation{\IFJ}
\author{H.~Clement}     \affiliation{\PITue}\affiliation{\Kepler}
\author{E.~Czerwi{\'n}ski}\affiliation{\IPJ}
\author{K.~Demmich}     \affiliation{\MS}
\author{R.~Engels}      \affiliation{\IKPJ}
\author{A.~Erven}       \affiliation{\ZELJ}
\author{W.~Erven}       \affiliation{\ZELJ}
\author{W.~Eyrich}      \affiliation{\Erl}
\author{P.~Fedorets}    \affiliation{\IKPJ}\affiliation{\ITEP}
\author{K.~F\"ohl}      \affiliation{\Giess}
\author{K.~Fransson}    \affiliation{\IKPUU}
\author{F.~Goldenbaum}  \affiliation{\IKPJ}
\author{A.~Goswami}     \affiliation{\IKPJ}\affiliation{\IITI}
\author{K.~Grigoryev}   \affiliation{\IKPJ}\affiliation{\HepGat}
\author{L.~Heijkenskj\"old}\altaffiliation[present address: ]{\Mainz}\affiliation{\IKPUU}
\author{V.~Hejny}       \affiliation{\IKPJ}
\author{N.~H\"usken}    \affiliation{\MS}
\author{L.~Jarczyk}     \affiliation{\IPJ}
\author{T.~Johansson}   \affiliation{\IKPUU}
\author{B.~Kamys}       \affiliation{\IPJ}
\author{G.~Kemmerling}\altaffiliation[present address: ]{\JCNS}\affiliation{\ZELJ}
\author{A.~Khoukaz}     \affiliation{\MS}
\author{A.~Khreptak}    \affiliation{\IPJ}
\author{D.A.~Kirillov}  \affiliation{\HeJINR}
\author{S.~Kistryn}     \affiliation{\IPJ}
\author{H.~Kleines}\altaffiliation[present address: ]{\JCNS}\affiliation{\ZELJ}
\author{B.~K{\l}os}     \affiliation{\Katow}
\author{W.~Krzemie{\'n}}\affiliation{\ASWarsH}
\author{P.~Kulessa}     \affiliation{\IFJ}
\author{A.~Kup\'{s}\'{c}}\affiliation{\IKPUU}\affiliation{\ASWarsH}
\author{K.~Lalwani}     \affiliation{\NITJ}
\author{D.~Lersch}\altaffiliation[present address: ]{\FSU}\affiliation{\IKPJ}
\author{B.~Lorentz}     \affiliation{\IKPJ}
\author{A.~Magiera}     \affiliation{\IPJ}
\author{R.~Maier}       \affiliation{\IKPJ}\affiliation{\JARA}
\author{P.~Marciniewski}\affiliation{\IKPUU}
\author{B.~Maria{\'n}ski}\affiliation{\ASWarsN}
\author{H.--P.~Morsch}  \affiliation{\ASWarsN}
\author{P.~Moskal}      \affiliation{\IPJ}
\author{W.~Parol}       \affiliation{\IFJ}
\author{E.~Perez del Rio}\altaffiliation[present address: ]{\INFN}\affiliation{\PITue}\affiliation{\Kepler}
\author{N.M.~Piskunov}  \affiliation{\HeJINR}
\author{D.~Prasuhn}     \affiliation{\IKPJ}
\author{D.~Pszczel}     \affiliation{\IKPUU}\affiliation{\ASWarsH}
\author{K.~Pysz}        \affiliation{\IFJ}
\author{J.~Ritman}\affiliation{\IKPJ}\affiliation{\JARA}\affiliation{\Bochum}
\author{A.~Roy}         \affiliation{\IITI}
\author{O.~Rundel}      \affiliation{\IPJ}
\author{S.~Sawant}      \affiliation{\IITB}
\author{S.~Schadmand}   \affiliation{\IKPJ}
\author{T.~Sefzick}     \affiliation{\IKPJ}
\author{V.~Serdyuk}     \affiliation{\IKPJ}
\author{B.~Shwartz}     \affiliation{\Budker}\affiliation{\Novosib}
\author{T.~Skorodko}\affiliation{\PITue}\affiliation{\Kepler}\affiliation{\Tomsk}
\author{M.~Skurzok}     \altaffiliation[present address: ]{\INFN}\affiliation{\IPJ}
\author{J.~Smyrski}     \affiliation{\IPJ}
\author{V.~Sopov}       \affiliation{\ITEP}
\author{R.~Stassen}     \affiliation{\IKPJ}
\author{J.~Stepaniak}   \affiliation{\ASWarsH}
\author{E.~Stephan}     \affiliation{\Katow}
\author{G.~Sterzenbach} \affiliation{\IKPJ}
\author{H.~Stockhorst}  \affiliation{\IKPJ}
\author{H.~Str\"oher}   \affiliation{\IKPJ}\affiliation{\JARA}
\author{A.~Szczurek}    \affiliation{\IFJ}
\author{A.~Trzci{\'n}ski}\altaffiliation{deceased}\affiliation{\ASWarsN}
\author{M.~Wolke}       \affiliation{\IKPUU}
\author{A.~Wro{\'n}ska} \affiliation{\IPJ}
\author{P.~W\"ustner}   \affiliation{\ZELJ}
\author{A.~Yamamoto}    \affiliation{\KEK}
\author{J.~Zabierowski} \affiliation{\ASLodz}
\author{M.J.~Zieli{\'n}ski}\affiliation{\IPJ}
\author{J.~Z{\l}oma{\'n}czuk}\affiliation{\IKPUU}
\author{P.~{\.Z}upra{\'n}ski}\affiliation{\ASWarsN}
\author{M.~{\.Z}urek}   \altaffiliation[present address: ]{\LBL}\affiliation{\IKPJ}

\newcommand*{\Mainz}{Institut f\"ur Kernphysik, Johannes 
 Gutenberg Universit\"at Mainz, Johann--Joachim--Becher Weg~45, 55128 Mainz, 
 Germany}
\newcommand*{\JCNS}{J\"ulich Centre for Neutron Science JCNS, 
 Forschungszentrum J\"ulich, 52425 J\"ulich, Germany}
\newcommand*{\FSU}{Department of Physics, Florida State University,
  77~Chieftan Way, Tallahassee, FL~32306-4350, USA}
\newcommand*{\INFN}{INFN, Laboratori Nazionali di Frascati, Via E. Fermi, 40, 
 00044 Frascati (Roma), Italy}
\newcommand*{\LBL}{Nuclear Science Division, Lawrence Berkeley Laboratory,
One Cyclotron Road, Berkeley, CA 94720-8153, United States}

\collaboration{WASA-at-COSY Collaboration}\noaffiliation

\author{A.~Deltuva}
\affiliation{Institute of Theoretical Physics and Astronomy, Vilnius University, Saulėtekio al. 3, LT-10257 Vilnius, Lithuania}
 \author{J.~Golak}
 \affiliation{Institute of Physics, Jagiellonian University, prof.\ 
 Stanis{\l}awa {\L}ojasiewicza~11, 30-348 Krak\'{o}w, Poland}
\author{A. Kozela}
\affiliation{The Henryk Niewodnicza{\'n}ski Institute of Nuclear Physics, 
 Polish Academy of Sciences, Radzikowskiego~152, 31--342 Krak\'{o}w, Poland}
 \author{R.~Skibi\'nski}
\affiliation{Institute of Physics, Jagiellonian University, prof.\ 
 Stanis{\l}awa {\L}ojasiewicza~11, 30-348 Krak\'{o}w, Poland}
\author{I.~Skwira-Chalot}
\affiliation{Faculty of Physics, University of Warsaw, Warsaw, Poland}
\author{H.~Wita\l a}
\affiliation{Institute of Physics, Jagiellonian University, prof.\ 
 Stanis{\l}awa {\L}ojasiewicza~11, 30-348 Krak\'{o}w, Poland}




\date{\today}

\begin{abstract}
Differential cross section for the $^{1}$H$(d,pp)n$ breakup reaction at deuteron beam energy 
of 340 MeV has been measured with the use of WASA detector at COSY-J\"{u}lich. The set of 
proton-proton coincidences registered at Forward Detector has been analysed on dense grid of
 kinematic variables, giving in total around 5600 data points.  
 The cross section data are compared to theoretical predictions based on the state-of-the-art
  nucleon-nucleon 
  potentials, combined with three-nucleon force, Coulomb interaction or carried out in a 
  relativistic regime. 
\end{abstract}


\maketitle 

\section{\label{secI}Introduction}

Properties of few-nucleon systems at medium energies are determined to large extent by
pairwise nucleon-nucleon (NN) interaction, which are a dominant component of the nuclear
 potential. NN interactions are described either by the realistic potentials \cite{Mac01, Wir95, Sto94}~or the potentials  derived from Chiral Effective Field Theory (ChEFT) \cite{Bed02, Wei90, Epe09}, achieving in both cases  
a  precise  description of  observables for two-nucleon systems. 
The deficiencies in description of systems consisting of three and more nucleons
are usually attributed to additional part of dynamics, beyond the NN interactions. The 
so-called three-nucleon force (3NF) is interpreted as a consequence of internal degrees of
 freedom of interacting nucleons. The 3NF arises  in the  meson-exchange picture as an intermediate excitation of a nucleon 
to a $\Delta$ isobar.  State-of-the-art models of 3NF's, like 
TM99 ~\cite{Coo01},  Urbana~IX~\cite{Pud97}, or Illinois \cite{Pie01},  
combined with the realistic  nucleon-nucleon (2N)  potentials, constitute the basis for  calculations of binding energies and scattering observables.   Chiral  Effective Field Theory provides a systematic 
construction of nuclear forces  in a fully consistent way: the 3N forces
 appear naturally at a certain order \cite{Epe09,Mac16}. The  theoretical   calculations including semi-phenomenological 3NF or 3NF stemming from 
 ChEFT,  reproduce with high accuracy binding energies of light 
 nuclei \cite{Viv98, Nog00, Wir01, Nav07}.  They provide also significantly improved description of differential 
 cross section for elastic nucleon-deuteron  scattering as compared to the calculations 
 based on NN interactions only \cite{Wit98, Hat02, Mer04}. Improvement in the sector of polarization observables is not 
 so clear, see discussion in Refs.\cite{Kal12, Kis13}, but this issue  will not be further 
 considered in this paper which is focused on the cross section.   
 However, even in the sector of the differential cross section, at beam energies above 100
  MeV/nucleon 
certain discrepancies between  the scattering data and calculations persist. 
Neither  Coulomb interactions between protons~\cite{Del05} nor relativistic 
effects~\cite{Wit05} are able to explain that observation, since their impact, except for very forward angles, where the Coulomb interaction plays a decisive role, is very small in this energy range. 

Studies of the $^{1}$H$(d,pp)n$ and $^{2}$H$(p,pp)n$ breakup reactions make important 
contribution to investigations of the 3NF effects. The advantage relies on kinematic 
richness of the three-body  final state. There are experimental evidences of 
significant 3NF contributions 
to the differential cross section for the breakup reaction, starting at relatively 
low beam energy of 65 MeV/nucleon \cite{Kis05, Kis13}. In contrast to the elastic scattering, 
Coulomb interaction  is a very important component of the breakup reaction dynamics. 
The Coulomb interaction between protons modifies  the cross section data  over 
 significant part of the phase space, in particular at forward laboratory 
angles of the $^{1}$H$(d,pp)n$ reaction \cite{Kis06, Cie15}. 
The Coulomb effects are dominating  in the region of configurations characterised 
with low relative momentum 
of the proton pair, the so called  proton-proton Final State Interaction (FSI)
 configurations.    At present, the Coulomb interaction and 3N forces are both included 
  into theoretical calculations and their interplay  can be studied \cite{Del05a, Del06, Del09}. 
  
   At energies  above 140 MeV/nucleon, practically there are no data for the breakup reaction. The only exception, measurement at 190\,MeV~\cite{Mar08, Mar08a}, provided  hints of
  deficiencies in description of   
the cross section for the deuteron breakup reaction, even when 3NF is included.  
 The problem can be interpreted either as confirmation of mentioned earlier problems 
 observed  in elastic scattering cross section,  or as a consequence of relativistic effects. 
 In contrary to elastic scattering,  
 the relativistic calculations of the differential cross section for  breakup reaction 
 lead   to different results  than the non-relativistic ones \cite{Wit11}.  Due to the 
 significant predicted 3NF and 
  relativistic effects in the energy region between 150 and 200 MeV/nucleon \cite{Ski06, Wit11}, the question  
  arises about their interplay. So far, there has been  no calculation with 
  full relativistic treatment of NN and 3NF interactions. Under such circumstances one has to
   rely on  systematic (in beam energy) studies over large phase space regions, with the 
   aim to single out both   contributions on the basis of their different 
   kinematic dependencies. 
   
   An experiment to investigate the $^{1}$H$(d,pp)n$ breakup reaction   
    using a~deu\-teron beam of 300, 340, 380 and 400\,MeV (150, 170, 190, 200\,MeV/nucleon) and the WASA (Wide Angle Shower Apparatus) detector, has been  
    performed at the Cooler Synchrotron COSY-J\"{u}lich.
 Due to almost $4\pi$ acceptance and moderate detection 
 threshold of the WASA system,  differential cross  section data have been collected in
  a large part of  the breakup reaction  phase space. As a~first step  the data collected at the beam 
  energy of 170\,MeV/nucleon have been analysed, with a focus on the proton-proton coincidences
   registered 
  in the Forward Detector.    

\section{\label{secII}Experiment and data analysis}

\subsection{\label{secIIA}Setup and measurement procedure}

\begin{figure}[h]
\centering
\includegraphics[width=85mm]{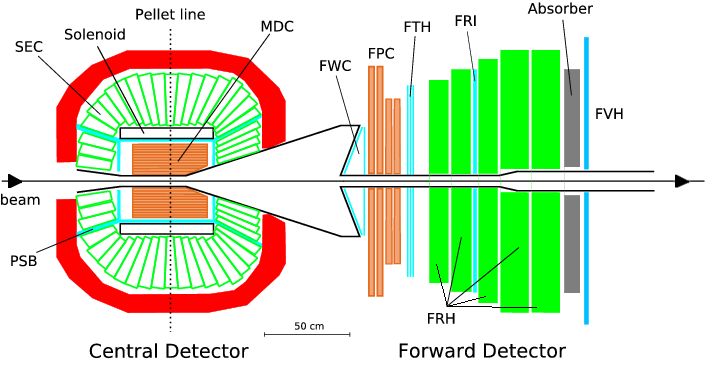}
\caption{(Color online) Schematic view of the detection system. }
\label{det}      
\end{figure}

 The WASA  detector \cite{Bar08, Ada04, Pod11}, covering almost full solid angle, consists of four main components: Central Detector (CD), Forward
Detector (FD), Pellet Target Device and  Scattering Chamber (see Fig.~\ref{det}).

COSY has functionality to group up different machine settings
within a ``supercycle'' which allows to change the beam energy in discrete steps from cycle to cycle \cite{Bec99}. This feature is very useful for the purpose of comparing the cross section at various beam energies.  
During the $^{1}$H$(d,pp)n$  measurement energies 170, 190 and 200\,MeV/nucleon of the
 deuteron beam were changed in supercycle mode of time length 30\,s (measurement at 150\,MeV/nucleon  
 was performed separately), using a barrier bucket cavity.  A barrier bucket cavity can be 
 used  to compensate the beam energy loss induced by an internal pellet target \cite{Sto10}. 

The pellet target is a unique development for the CELSIUS/WASA 
experiment.  The target provides a narrow stream of very small frozen hydrogen or deuterium 
droplets with diameters down to 25\,$\mu$m, called pellets. 
Some of the parameters of the pellet target are listed in Table~\ref{tab-pellet}.

\begin{table}[h]
\centering
\caption{Performance of the pellet target system \cite{Bar08}}
\label{tab-pellet}     

\vspace{0.2cm}

\begin{tabular}{ll}
\hline
\hline
pellet diameter ($\mu $m)  & 25-35  \\
pellet frequency (kHz) & 5-12 \\
pellet-pellet distance (mm) & 9-20  \\ 
beam diameter (mm) & 2-4\\
effective target area density (atoms/cm$^2$) &   $>$10$^{15}$\\
\hline
\hline
\end{tabular}
\end{table}


FD covers the region of the polar angles from 
2.5$^{\circ}$ to 18$^{\circ}$. It consists of a set of detectors 
for the identification of charged hadrons and track reconstruction:
Forward Window Counter (FWC), Forward Proportional Chamber (FPC), Forward Trigger 
 Hodoscope (FTH), Forward Range Hodoscope (FRH) and Forward Veto Hodoscope (FVH). 
 Between the second and third layers of FRH there are two  layers of Forward Range
  Interleaving  Hodoscope (FRI). FPC is used for precise determination of 
 particle emission angles. The FD plastic scintillators are used for particle 
 identification and particle energy measurement. They all provide information for the first level trigger logic. Some features of the FD are given in Table~\ref{tab-fd}. CD was used in the experiment described here, but the present data analysis do not include particles registered in that part. For description of the CD see Refs.~\cite{Bar08, Ada04}.

\begin{table}[h]

\caption{ Basic information on the FD}
\label{tab-fd}     

\vspace{0.2cm}

\centering
\begin{tabular}{ll}
\hline
\hline
number of scintillator elements   & 340  \\
scattering angle coverage  & 2.5$^{\circ}$-18$^{\circ}$ \\
scattering angle resolution & 0.2$^{\circ}$ \\
 amount of sensitive material & 50\,g/cm$^2$\\
\hspace{0.2cm} -~in radiation length &   $\approx$ 1\,g/cm$^2$\\
\hspace{0.2cm} -~in nuclear interaction  length &   $\approx$ 0.6\,g/cm$^2$\\
maximum kinetic energy (T$_{stop}$)  & \\
for stopping $\pi ^{\pm}$/p/d/$^4$He & 170/340/400/900\,MeV\\
time resolution & $\le$ 3\,ns \\
energy resolution for & \\
\hspace{0.2cm} stopped particles & 1.5\%-3\% \\
\hspace{0.2cm} particles with T$_{stop}<$ T $<$ 2T$_{stop}$   & 3\%-8\% \\
particle identification  & $\Delta$E-E, $\Delta$E-$\Delta$E \\
\hline
\hline
\end{tabular}
\end{table}

During data taking for the dp breakup experiment described here there were a few trigger types 
in use. Trigger named No.~7  was the 
basic trigger for registering events in FD detector. It required at least one track with 
correct matching of clusters in FWC, FTH and FRH. It was later used in the analysis of single tracks of deuterons from the elastic scattering and of proton-proton coincidences from the breakup reaction. Trigger named No.~2 was much less restrictive: one hit above threshold was required. Due to high rate of events accepted by this trigger prescaling by a factor 10 was necessary. The comparison of results obtained with triggers 2 and 7 is important for controlling possible bias imposed by trigger conditions.

\subsection{\label{secIIB}Data analysis}

The data analysis presented in this work is focused on the proton-proton coincidences  from the $^{1}$H$(d,pp)n$  breakup reaction at 170\,MeV/nucleon  registered  in the FD. The  aim of our study is the determination of  the differential cross section on a dense angular grid of kinematical configurations defined by the emission angles of the two outgoing protons: two polar angles $\theta_1$ and $\theta_2$
(in the range between 5$^{\circ}$ and 15$^{\circ}$) and the relative azimuthal angle $\varphi _{12}$ (in the wide range between 20$^{\circ}$ and 180$^{\circ}$).

\subsubsection{Event selection and particle identification}
\begin{figure}[h]
\begin{center}
\subfigure{
\includegraphics[width=7cm]{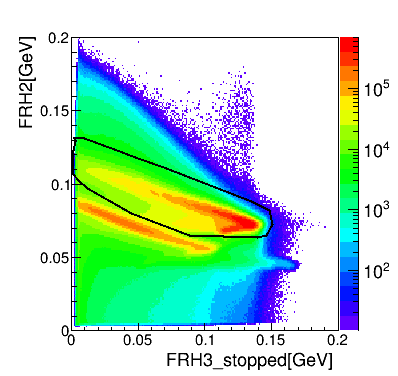}
}
\subfigure{
\includegraphics[width=7cm]{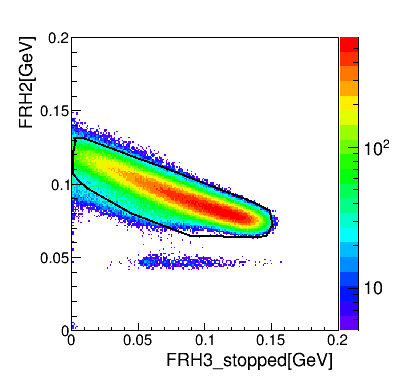}
}\\
\end{center}

\vspace{-0.7cm}

\caption{Particle identification spectra for particles stopped in the 3rd layer of FRH. The ``banana-shaped'' gate represents the cut applied to select deuterons, the same for experimental data (top panel) and Monte Carlo simulation (bottom panel). }
\label{fig-pid}      
\end{figure}

The first step of data analysis is the selection of events of interest, 
$\it i.e.$ two protons from the breakup process and deuterons from elastic scattering 
channel registered in the FD. The particle identification is based on the $\Delta E$-$E_R$ technique, where $E_R$ is remaining energy deposit in the layer where particle is stopped (see example in  Fig.~\ref{fig-pid}, top panel). In the whole range of energies,
a clear separation between loci of protons and deuterons is observed.
The analogous spectra are built for data generated in Monte Carlo simulation, separately
 for elastic scattering and breakup reaction, see example of the deuteron distribution in Fig.\ref{fig-pid}, bottom panel. The simulation is used to verify cuts set on the data. The only difference between
  experimental data and simulation spectra is due to particles
 punching-through the 3rd layer of FRH and stopped in the inactive layer behind it (not included in the
  simulation). For those events total energy is reconstructed on the basis of energy loss in the 2nd
  layer.

\subsubsection{Energy calibration}

Energy calibration of FD is based on measurements of dp elastic scattering at 
energies corresponding to minimum ionization with non-uniformity and nonlinearity corrections,  as described in detail in Ref.~\cite{Vla08}. Since the FRI detector was not used in a number of 
previous runs, its calibration is not included in the main calibration procedure and is 
known with lower accuracy. The appropriate corrections have been applied, but  
in case of  protons stopped in FRI (protons with initial energy of about 200 MeV) the energy resolution is diminished. In order to avoid the systematic error related to this effect, the affected energy region  has been rejected from the cross section analysis.

\subsubsection{Analysis of the breakup reaction}
 \label{anal-breakup}

\begin{figure}[h]
\begin{center}
\includegraphics[width=7cm]{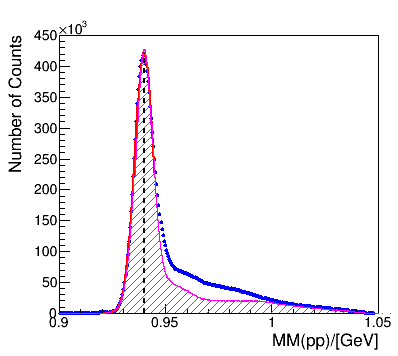}
\end{center}

\vspace{-0.5cm}

\caption{(Color online) Missing mass reconstructed from momenta of two outgoing protons detected in coincidence (full points). Clearly, the peak corresponding to the neutron mass dominates. Red line represents Gaussian fit with mean value of 0.94\,GeV. 
The right tail of the distribution originates from proton energy loss due to hadronic 
interactions and from accidental coincidences. The data are compared to histogram built on the basis of MC simulation (hatched magenta histogram).}
\label{miss}     
\end{figure}

The missing mass spectrum is a tool to control the proton energy calibration and the procedure of selection of proton-proton coincidences. The missing mass of the neutron is calculated according to the formula (in which $c=1$):

\begin{equation}
MM = \sqrt{(E_{in}-E_{p_1}-E_{p_2})^{2}- (\vec{P}_{in}-\vec{P}_{p_1}-\vec{P}_{p_2})^{2}},
 \label{mm}
\end{equation}

\noindent
where $E_{in}$ and $\vec{P}_{in}$ are the sum of energy and momenta of the incident deuteron and target proton and 
$E_{p_i}$ and $\vec{P}_{p_i}$ ($i=1,2$) are the total energies and momenta of the two outgoing protons registered in coincidence. Fig.~\ref{miss} presents the missing mass spectrum, 
built for all pairs of coincident  protons registered in FD.
Similar histogram has been built for breakup events generated with Monte Carlo simulation. Since all the
cuts applied in analysis procedures are the same for experimental and simulated data,
the model of hadronic interactions applied in simulation can be verified by comparing the missing mass
spectra. This check is in turn important for efficiency corrections. The qualitative agreement of shapes can be observed, while the
 remaining differences can be attributed to background of accidental coincidences and influence of
 electronic thresholds on the data, both mechanisms absent in the simulation. The missing mass histograms for data and MC will be further discussed, also quantitatively, in sec.~\ref{consist}.

After the selection of proton-proton coincidences and having performed the energy calibration,  any kinematical configuration of the breakup reaction within the angular acceptance of the detection system can be analysed. The configuration has been defined by emission angles of the two outgoing protons: two polar angles $\theta _1$ and $\theta _2$  and their relative azimuthal angle $\varphi _{12}$.
 The data are integrated over the angular ranges of  $\theta _{1,2}$ ($\pm $1$^{\circ}$) and $\varphi _{12}$ ($\pm$ 5$^{\circ}$). These ranges are large as compared to angular resolution of the detectors and, therefore, no significant systematic uncertainty is related to the determination of solid angles obtained for selected configurations. The effect of averaging of cross section within the angular ranges is taken into account when comparing the data with the theoretical calculations, which have been 
 averaged accordingly 
and projected onto relativistic kinematics  \cite{Kis13}

\begin{figure}[ht]
\begin{center}
\subfigure{
\includegraphics[width=6.5cm]{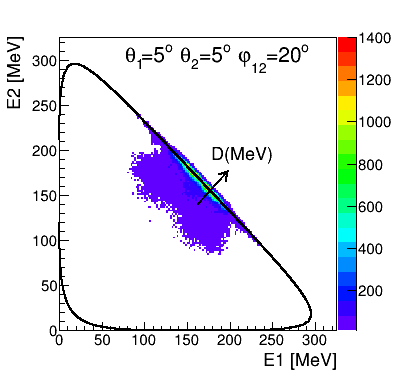}
}
\subfigure{
\includegraphics[width=6.5cm]{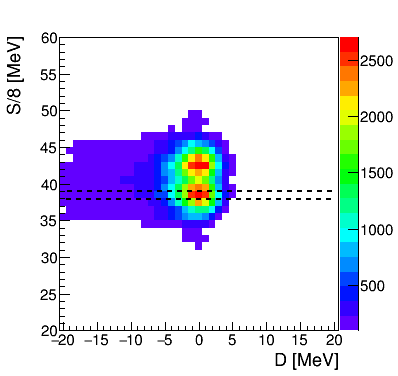}
}\\
\end{center}

\vspace{-0.5cm}

\caption{(Color online) {\it Top panel} : $E_1$ vs. $E_2$ coincidence spectrum of the two protons registered at $\theta _1$=5$^{\circ}\pm$ 1$^{\circ}$, $\theta _2$=5$^{\circ}\pm$ 1$^{\circ}$, and $\varphi _{12}$=20$^{\circ}\pm$ 5$^{\circ}$. The solid line shows a three-body kinematical curve, calculated for the central values of experimental angular ranges. $D$ axis illustrates the distance of  the ($E_1$, $E_2$) point from the kinematical curve. {\it Bottom panel}: transformation of $E_1$ vs. $E_2$ spectrum to $S$ (arclength) vs. $D$ (distance from kinematical line in $E_1$-$E_2$ plane). Dashed lines represent integration limits ($\Delta S$=8\,MeV) for a sample $S$-slice.  }
\label{kin-svsdd}      
\end{figure}

The sample kinematical spectrum $E_1$ versus~$E_2$ obtained for selected configuration 
is shown in Fig.~\ref{kin-svsdd}, top panel. The 
center of the band formed by experimental data is lying on the relativistic kinematics curve (corresponding to the point-like, central geometry). Correct kinematic relations of the data  confirm accuracy of the energy calibration. In the next step, new variables are introduced: $D$ is the distance of the ($E_1$,~$E_2$) point from the kinematic curve in the $E_1$-$E_2$ plane and $S$ denotes 
 the value of the arclength along the kinematical line  with the starting point ($S$=0) chosen arbitrarily at the point  where $E_2$=~0 and starts to rise.  
The events contained within the 
distance $D$ of $\pm$20\,MeV from the kinematical line are selected for further analysis and presented in  $S$ vs.~$D$ spectrum (see Fig.~\ref{kin-svsdd}, bottom panel).

\begin{figure}[h]
\begin{center}
\subfigure{
\includegraphics[width=7cm]{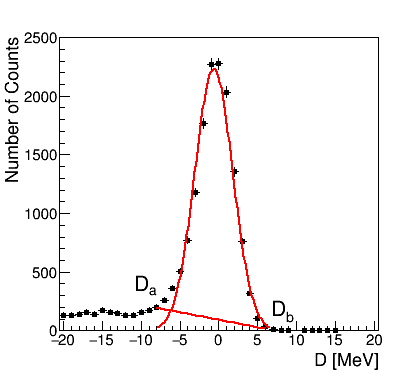}
}
\subfigure{
\includegraphics[width=7cm]{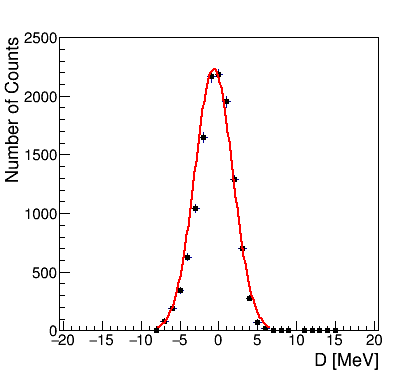}
}\\
\end{center}

\vspace{-0.5cm}

\caption{(Color online) {\it Top panel}: Determination of the background contribution in  one slice  in the $S$ vs.~$D$ spectrum (Fig. \ref{kin-svsdd}, bottom panel). The background is estimated by a linear function between  limits of integration ($D_a$, $D_b$) (shown with the solid red line). {\it Bottom  panel}~:  $D$-projected distribution after the background subtraction with  a Gaussian distribution fitted in the range of $D$ corresponding to distance of -3$\sigma$ and +3$\sigma$ from the peak center. }
\label{subt}      
\end{figure}

\begin{figure}[h]
\begin{center}
\includegraphics[width=8.5cm]{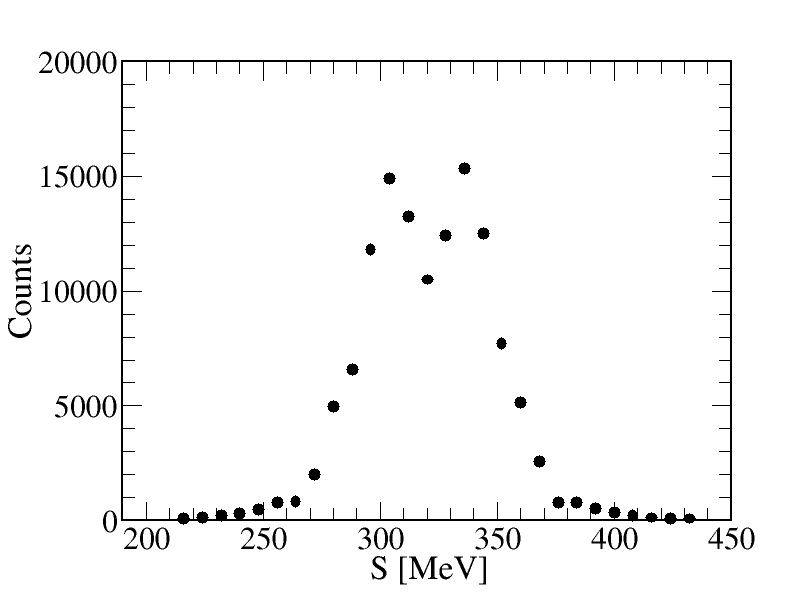}
\end{center}

\vspace{-0.5cm}

\caption{An example of  $S$ distribution of the  rate  of breakup events
  obtained for the chosen kinematical configuration ($\theta _1$=5$^{\circ}\pm$1$^{\circ}$, $\theta _2$=5$^{\circ}\pm$ 1$^{\circ}$, and $\varphi _{12}$=20$^{\circ}\pm$ 5$^{\circ}$).
 Statistical errors are smaller than the point size. }
\label{count_break}      
\end{figure}

The procedure of  background subtraction  is presented in Fig.~\ref{subt}. 
Each slice on the $S$ vs.~$D$ spectrum (see Fig.~\ref{kin-svsdd}, bottom panel) is treated 
separately.  The background is approximated by a linear function between the two limits of integration ($D_a$, $D_b$) defined as -3$\sigma$ and +3$\sigma$ from the peak position (Fig.~\ref{subt}, top panel). 
The $D$-projected distributions obtained after the background subtraction have Gaussian shape 
(with exception of bins characterised by low signal-to-background ratio, see discussion 
in Sec.~\ref{error}). 
The Gaussian distribution is fitted in the range from  $D_a$ to $D_b$ (see Fig.~\ref{subt}, bottom panel).  Number of events obtained after background subtraction  is presented   as a function of the arclength $S$, see an example  in Fig.\,\ref{count_break}. After normalization to the integrated luminosity, the differential cross section is~obtained.

\subsubsection{Cross section normalization \label{normaliz}}

For the purpose of normalization of the experimental results, the luminosity is determined on the basis of the number of the elastically-scattered deuterons.

\begin{figure}[h]
\begin{center}
\includegraphics[width=7cm]{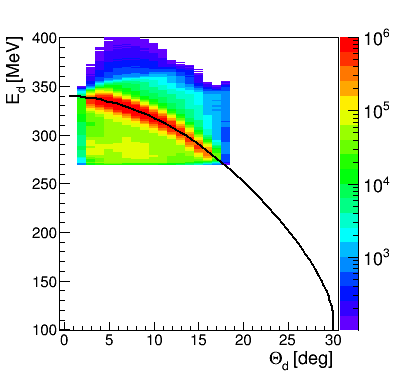}
\end{center}

\vspace{-0.5cm}

\caption{ (Color online) The energy vs.  $\theta_d$ polar scattering angle distribution for events identified as deuterons. Solid curve corresponds to kinematics of elastically scattered deuterons. }
\label{fig-prost}      
\end{figure}

Selection of deuterons registered in the FD has been based on the $\Delta E$-$E_R$ technique.  After applying  energy calibration for protons the energy calibration for deuterons has been readjusted with the use of  MC simulation.  Kinematics of deuterons registered in FD obtained after the  corrections is shown  in Fig.~\ref{fig-prost}. Particles which have not reached the 3rd FRH layer are not accepted.

\begin{figure}[h]
\begin{center}
\includegraphics[width=8.5cm]{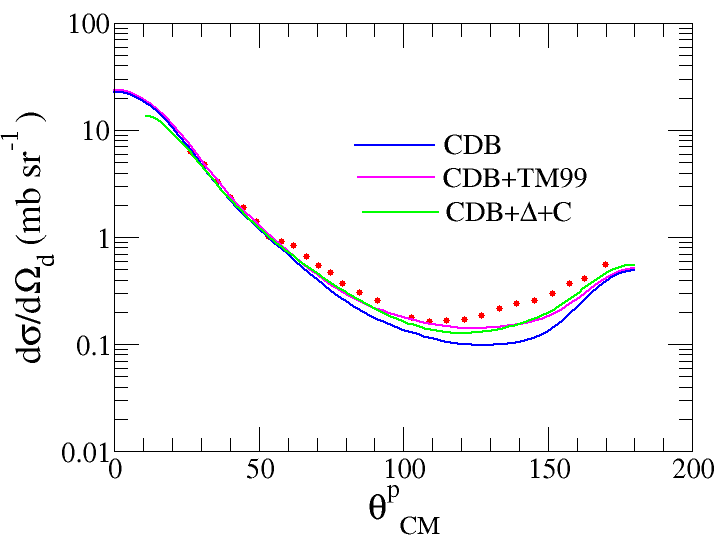}
\end{center}

\vspace{-0.5cm}

\caption{(Color online) Angular distribution of the elastic scattering cross section  in the CM 
system. Red dots represent the measured cross-section values for elastic scattering at E$_p$=170 MeV\cite{Erm03}. The solid lines show the results of the theoretical calculations with the CD Bonn potential and the TM99 3NF as well as coupled-channel potential CD Bonn+$\Delta$ and Coulomb force included (CDB+$\Delta$+C). }
\label{fig-elast340}      
\end{figure}

\begin{figure}[h]
\begin{center}
\includegraphics[width=8.5cm]{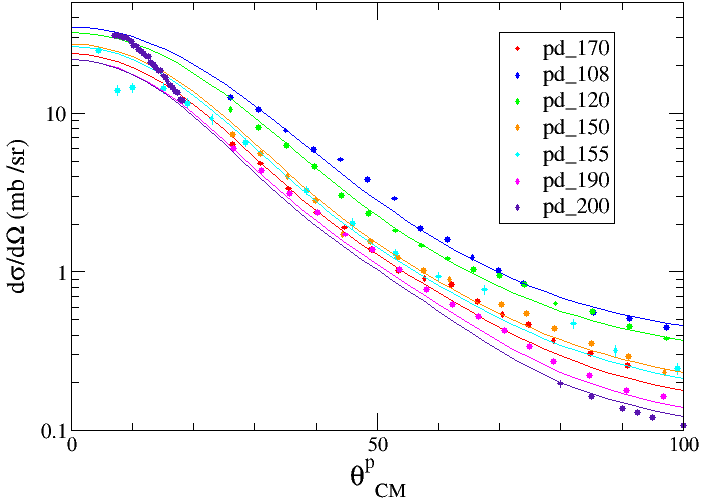}
\end{center}

\vspace{-0.5cm}

\caption{(Color online) Experimental differential cross section of the reaction $^2$H$(p,dp)$ in the CM system in the angular range of FD,  at the incident-beam energies: 108\,MeV, 120\,MeV, 150\,MeV, 170\,MeV, 190\,MeV \cite{Erm03} and 155\,MeV \cite{Kur64}, 200\,MeV  \cite{Ade72, Roh98}. The solid lines show the results of the theoretical calculations  with the CD Bonn potential and the TM99 3NF.} 
\label{exp_teoria_zoom}      
\end{figure}

\begin{figure}[h]
\begin{center}
\includegraphics[width=6.5cm]{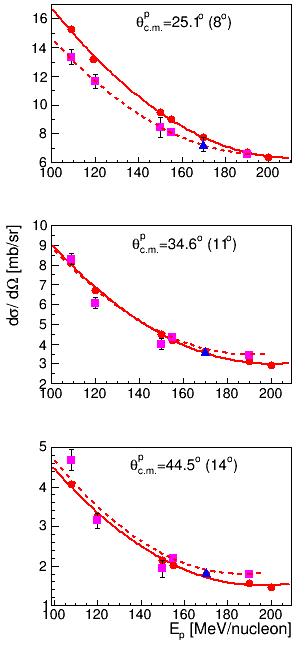}
\end{center}

\vspace{-0.5cm}

\caption{(Color online) Experimental (triangles, squares and dashed lines) and theoretical (full points and 
solid lines)  cross section for elastic scattering at given $\theta_{c.m.}$ angle (corresponding $\theta_{lab}$ in the bracket)  presented in
 function of incident beam energy per nucleon. Dashed lines represent polynomial fitted to experimental points (excluding the point at 170\,MeV). The solid  lines present fitted functions to points obtained from theoretical calculations. }
\label{fig-teoria_theta_3}      
\end{figure}

In order to obtain the luminosity  a reference cross section $\sigma_{LAB}^{el}$ for elastic scattering  at the studied energy should be known at angle contained within the acceptance of our detector.  It is the case, although the $^2$H$(p,dp)$ cross section distribution measured at 170\,MeV  reveals irregularities which suggest systematic errors, see Fig.~\ref{fig-elast340}.  It is extremely difficult to control the absolute cross section value with an accuracy of 5\%: The comparison of experimental data with theoretical calculation including the 3NF, \cite{Erm03, Kur64, Sek02, Ade72, Roh98} has shown  not only rising with energy deficiency of calculations at the cross section minimum, but also scatter of the data  exceeding their statistical errors~\cite{Ahm09}. In order to minimize the bias of the results, the normalization is based on all the data sets from the range of energies (between 108 and 200\,MeV) compared to theoretical predictions. 
Deuterons scattered at angles covered by FD correspond to  $\theta^{p}_{CM}<$50$^{\circ}$, where the theoretical calculations including 3NF provide precise description of the data, 
see Fig.~\ref{exp_teoria_zoom}. On the basis of the available data, the  dependence of cross
 section on beam energy can be studied at each polar angle, see examples in 
Fig.~\ref{fig-teoria_theta_3}. 
Theoretical calculations including TM99 3NF (full points, solid lines) provide consistent 
description of the data, with exception of  the lowest studied angle of 8$^\circ$ in the 
laboratory system. Trends of experimental data (polynomials presented as dashed lines) are  
based on all the data points (squares) but the one measured at 170\,MeV (triangle). Finally, we applied  three ways to obtain luminosity: taking  values of the  cross section $\sigma_{LAB}^{el}(\theta_d)$ given by (a) calculations, (b)  measurement at 170\,MeV,  and (c)  
 the polynomial fit to other data sets. In each case the following formula is 
 used to obtain 
the luminosity integrated over the measurement time:
\begin{equation}
L = \frac{N_{el}(\theta_d)}{\sigma_{LAB}^{el}(\theta_d) \Delta\Omega_d\epsilon^{el}(\theta_d)}, \label{lum}
\end{equation}

\begin{figure}[h]
\begin{center}
\includegraphics[width=7.5cm]{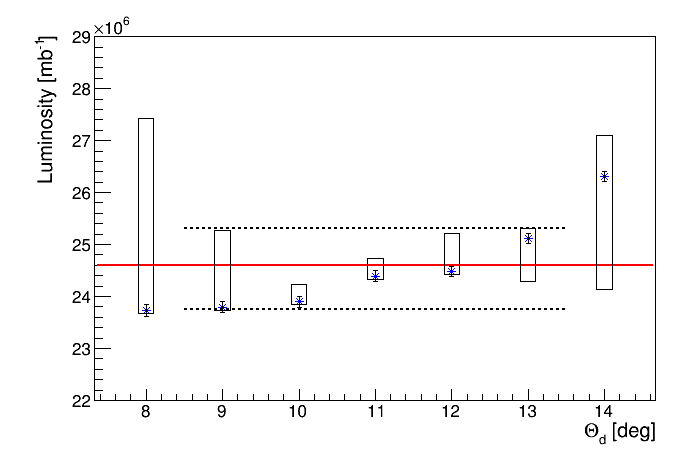}
\end{center}

\vspace{-0.6cm}

\caption{(Color online) Determination of the integrated luminosity. The values of the luminosity are presented as a  function of the deuteron scattering polar angle. The stars show results obtained with method (c).  The solid line corresponds to the weighted average of five results with the smallest systematic errors (shown as boxes). The dashed lines represent the range of systematic error of the averaged luminosity (2\%-3\%). }
\label{fig-luminosity}      
\end{figure}

\vspace{0.5cm}

\noindent
where $N_{el}$ is a number of elastically  scattered deuterons registered at the deuteron emission   
angle  $\theta_d$ (during the certain time), $\Delta\Omega_d$  is the solid angle for  
registering  deuterons and $\epsilon^{el}(\theta_d)$ is a detection efficiency for deuterons 
determined with the use of MC simulation.  In order to 
 control the result, the procedure is repeated for each deuteron polar angle between 
   8$^\circ$ and 14$^\circ$, see Fig.~\ref{fig-luminosity}. The spread of luminosity values 
   obtained at  the lowest deuteron polar angle of  8$^\circ$ is large, as expected from the 
   above discussion.   At the largest angles, 13$^\circ$ and 14$^\circ$,   significant
    systematic 
 uncertainty is related to proton background leaking through the deuteron 
 gate. The contribution of this background is estimated on the basis of MC simulation.  Conservatively, the largest error due to neglecting this contribution is taken as the systematic uncertainty. 
 The  ranges of  luminosity values  obtained at all studied angles are  consistent with 
 each other. The final result  has been obtained  neglecting points at marginal angles of 8 
 and 14 degrees,  due to their large systematic errors. 
Finally, the average integrated luminosity obtained for the full  set of data is (2.437 $\pm$ 0.005)~$\cdot$10$^{7}$mb$^{-1}$ with systematic error of -2\%, +3\%.
 It is presented in Fig.~\ref{fig-luminosity} as a solid horizontal   line with error limits shown as dashed lines.

The differential breakup cross section for a chosen angular configuration normalized to  
integrated luminosity value 
is given by the following formula:

\begin{equation}
\frac{d^5\sigma(S, \Omega_1, \Omega_2)}{d\Omega_1d\Omega_2dS} = \frac{N_{br}(S, \Omega_1, \Omega_2)}{L\,\Delta\Omega_1\Delta\Omega_2\Delta S\epsilon^{br}(S, \Omega_1, \Omega_2)}, \label{eq_break}
\end{equation}

\vspace{0.5cm}

\noindent
where $N_{br}$ is the number of breakup coincidences registered at the angles $\Omega_1$, $\Omega_2$ and projected onto a $\Delta S$-wide arclength bin. Subscripts 1 and 2 refer to two protons registered in coincidence.  Numbering of protons is defined by condition: $\theta_1 \leq \theta_2$ (for equal angles numbering is randomized).
$\Delta\Omega_i$, with i=1, 2, denotes the solid angles ($\Delta\Omega_i = 
\Delta\theta_i\Delta\varphi_i \, \sin\theta_i $) and  $\epsilon^{br}(S, \Omega_1, \Omega_2)$ 
is a product of all relevant 
efficiencies determined for each angular configuration.  
The normalization of the breakup cross section to the known cross section for elastic 
scattering, see Eq.~(\ref{lum}),  has the important advantage:  electronic dead-time, 
trigger efficiency,  the charge collected in Faraday Cup, the number of beam particles passing
 through the pellet target, etc.  
affect in the same way $N_{br}$ and $N_{el}$ and cancel in the ratio.

\begin{figure}[h]
\begin{center}
\includegraphics[width=8.6cm]{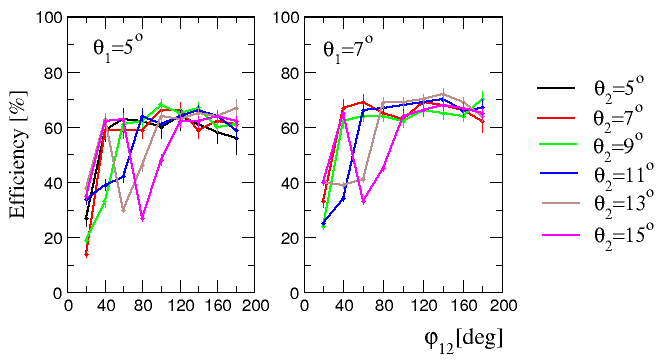}
\end{center}

\vspace{-0.5cm}

\caption{(Color online) Efficiency of the detection of proton-proton coincidences determined on the basis of MC simulation. The results for a number of selected kinematical 
configurations ($\theta _1$, $\theta _2$, $\varphi _{12}$) are shown as points with statistical errors. Lines are added to guide the eye.}
\label{br-eff}      
\end{figure}

WASA Monte Carlo program  was used for precise determination of efficiency of the detection system ($\epsilon$). Including detector acceptance and all cuts applied in the analysis, detector efficiency for registering and identifying elastically scattered deuterons is about 80\%. The efficiency of the detection system for proton-proton coincidences obtained for each kinematical configuration with defined integration limits: $\Delta\theta_1$=$\Delta\theta_2$=2$^{\circ}$ and $\Delta\varphi _{12}$=10$^{\circ}$ is presented in Fig.~\ref{br-eff}. The pits in the distributions reveal clear angular pattern, since they are caused by loss of events when both protons hit the same detector element. 
Due to low efficiency and possible inaccuracies related to the limit of the detection acceptance, configurations with $\theta_{1,2}$=17$^{\circ}$ were not included into analysis.

\subsubsection{Data consistency checks \label{consist}}

\begin{figure}[h]
\begin{center}
\includegraphics[width=8.5cm]{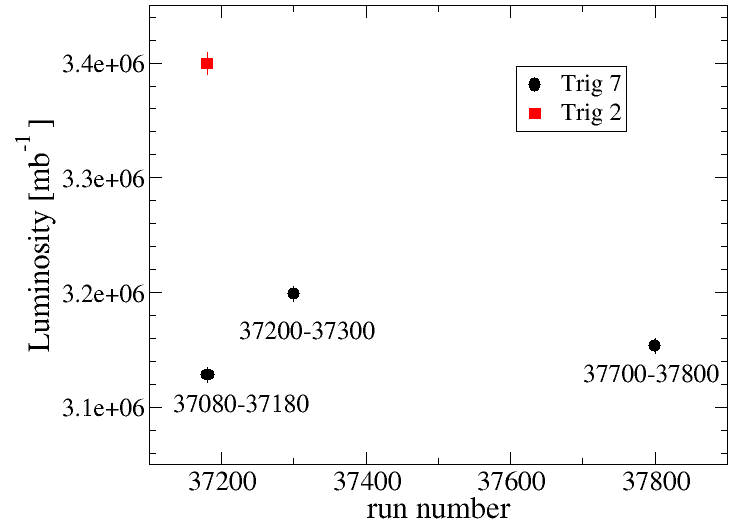}
\end{center}

\vspace{-0.7cm}

\caption{(Color online) Comparison of the luminosity values obtained  on the basis of data collected with different trigger and/or different time of the data collection (different ranges of the run numbers, as shown in the Figure). Statistical  errors are smaller than the point size.}
\label{lum-comp}      
\end{figure}

\begin{figure}[h]
\begin{center}
\includegraphics[width=8.5cm]{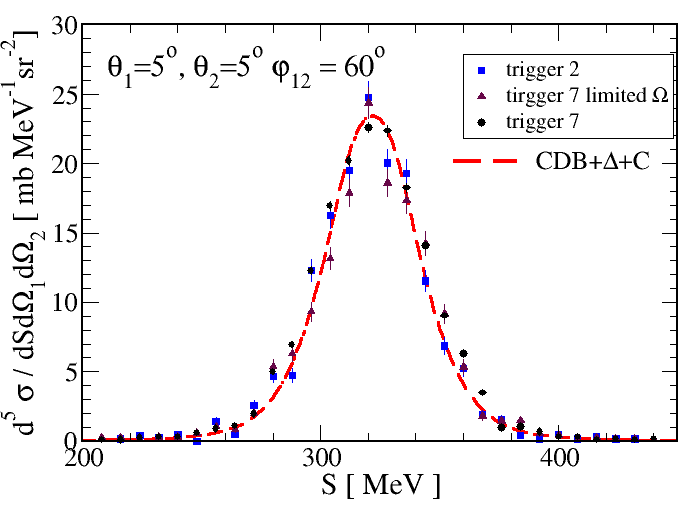}
\end{center}

\vspace{-0.6cm}

\caption{(Color online) Example of the differential cross section of the breakup reaction  as a function of 
the $S$ value obtained at different conditions: with triggers 2, 7 and  with trigger 7 
combined with narrower limits on the  
solid angle. The calculation within the coupled-channel potential CD Bonn+$\Delta$ 
 and the Coulomb force included  is represented by a red dashed  line. Full Widths 
at Half Maximum (FWHM) of the presented distributions are collected in Table \ref{fwhm}. }
\label{cross_comp}      
\end{figure}

The core analysis has been performed on the basis of the data collected with the main trigger in FD (trigger No.~7, see Sec.~\ref{secIIA}). 
In order to check consistency and stability of the result, the  luminosity is determined 
for three different data sets (of equal size) under condition of trigger No.~7 and
 for one 
of these data sets  under condition of much less restrictive trigger No.~2. 
The results (Fig.~\ref{lum-comp}) confirm stability of the obtained integrated luminosity
 values with the same trigger, while the difference between values of luminosity obtained for data collected with
  two trigger types is about 8\%. This might suggest  different  background 
  contribution to the events registered with these two triggers. However, it has been 
  checked that for both triggers   the background contribution is very similar, 
  of about 13\%-15\% (at angles $\theta_d <13^\circ$, where contribution of proton background is negligible).  On the other hand, 
  the same ratio of rates is obtained for the breakup data collected with those 
  triggers.  Therefore, we can interpret the difference as a loss of events due to the 
  restrictive trigger condition, i.e.~as an efficiency of the trigger. 
  In the next step the influence of the trigger on final 
  results for the breakup data is checked (see Fig.~\ref{cross_comp}). There is no
  statistically significant difference, which indicates that the 
  elastic scattering and breakup data are affected by the trigger efficiency in a similar
   way, which leads to cancellation of the effect in the ratio (Eq. \ref{eq_break}). 

\begin{table}[h]
\caption{Full Width at Half Maximum for distributions presented in Fig.~\ref{cross_comp}}\label{fwhm}
\vspace{-3mm}
\begin{center}
\begin{tabular}{|c|c|c|c|}
\hline
distributions  & Trigger & $\Delta\theta_1$,  $\Delta\theta_2$,  $\Delta\varphi_{12}$ & FWHM$\pm \Delta$FWHM \\
\hline
\hline
data & 7 & 2$^{\circ}$, 2$^{\circ}$, 10$^{\circ}$ & 60.0 $\pm$ 8.2 \\
\hline
data & 7 & 1$^{\circ}$, 1$^{\circ}$, 5$^{\circ}$ & 61.3 $\pm$ 8.1 \\
\hline
data & 2 & 2$^{\circ}$, 2$^{\circ}$, 10$^{\circ}$ & 59.6 $\pm$ 8.3 \\
\hline
CDB+$\Delta$+C & - & 2$^{\circ}$, 2$^{\circ}$, 10$^{\circ}$ & 56.3 \\
\hline
CDB+$\Delta$+C & - & only central values & 55.4 \\
\hline
\end{tabular}
\end{center}
\end{table}

It has been observed that data reveal systematically the wider $S$ distributions than all the
 theoretical predictions (see example in Table~\ref{fwhm}), even 
in spite of the fact that averaging over the angular ranges has been applied to the theoretical 
calculations. Nevertheless, the impact of averaging on the width of the distributions has been studied both in the data and calculations.  The comparison  of FWHM's shown in 
Table~\ref{fwhm} leads to the conclusion that the difference cannot be explained by effects 
of averaging. Procedure of projecting theoretical calculations onto the relativistic
 kinematics does not change the width of the resulting distribution, either. 
In Fig.~\ref{cross_comp} the cross section obtained with a limited solid angle  (the event 
integration ranges)  of  $\Delta\theta _1$=$\Delta\theta _2$=1$^\circ$ and $\Delta\varphi 
_{12}$=5$^\circ$ is compared to the one obtained in the standard analysis. There is no significant change in shape or height of the
 distributions.  The sums of data points (integrated distributions) are also equal within the limits of their statistical uncertainties.
 
 The efficiency corrections, discussed  in sec.~\ref{consist}, rely on simulations of particle interactions in the WASA detection system. Since hadronic interactions reduce registered energy of  particles, the neutron mass reconstructed from  momenta of two protons is distorted. Therefore the missing mass spectra are used to control this effect in the data and the MC simulations. The spectra presented in Fig.~\ref{miss} have been integrated in the region of +/- 3 sigma around the neutron peak (N$_{peak}$) and in the whole range (N$_{all}$) in order to compare the contribution of distorted events in the experimental and simulated data. The correction for the distorted events, N$_{all}$/N$_{peak}$ ratio, is 1.47 in experiment and 1.31 in MC simulations. The experimental data show relatively 12\% larger effect, but the difference can be partially attributed to contribution of accidental coincidences. From estimates of this contribution and taking into account partial cancellation of the hadronic effect in detection efficiency for deuterons and proton pairs (Eq.~\ref{mm}), systematic error of 7\% is attributed to possible underestimated hadronic interactions in Geant 3 simulation.

\subsubsection{Experimental uncertainties } \label{error}

Statistical errors of the measured cross-section values comprise an uncertainty  of the measured number of the breakup coincidences and of the luminosity. In all 189 kinematic  configurations 
the statistical error in maximum of the cross-section distribution is  2\% or less.

The systematic error of the cross-section stems primarily from three sources: detection 
efficiency, luminosity determination and background subtraction procedure. Uncertainty of
 determination of 
efficiency for proton-proton coincidences corresponds to statistical accuracy of MC 
simulation and  varies between 1\% and 4\%, reaching up to 
7\% for configurations with  the lowest $\varphi_{12}$=20$^{\circ}$ values  
(see Fig.~\ref{br-eff}). Typical uncertainty of the background subtraction is of about 5\%. 
Only at low cross section values, mainly at the tails of the studied $S$-distributions, 
 signal-to-background ratio reaches low values of 6 and  the
 background subtraction introduces significant uncertainty of 13\%. Systematic uncertainty 
 of the luminosity is between $-2$\% and $3$\%. 
 Finally, systematic uncertainties of data points vary between  6\% and 18\%, 
dominating the total uncertainty of the result. 
 They are presented as bands on the $S$ distributions (Figs.~\ref{fig5t5}~-~\ref{fig15t15}) and included into calculations of  $\chi^2$.

\section{Theoretical calculations} \label{theory}

Theoretical calculations for the systems of three nucleons are performed using exact nonrelativistic three-body theory. The system of coupled equations for transition operators,  proposed by Faddeev  \cite{Fad61} or Alt, Grassberger and Sandhas (AGS) \cite{Alt67},  are solved in momentum space. The models of nuclear interactions are the input to these calculations. Generally, the nuclear interactions applied to these calculations are constructed in one of  three different ways described below.
  
In the first approach,  semi-phenomenological models of nucleon-nucleon interaction are used,  
which base on the meson-exchange theory and have also a phenomenological part describing a short-range interaction with parameters 
 fitted to the two-nucleon data. There exist several so-called realistic NN potential models based on this approach, like  charge dependent (CD) Bonn  \cite{Mac01}, Argonne V$_{18}$ \cite{Wir95} and Nijmegen I and II
 \cite{Sto94}, providing an excellent  description of two-nucleon observables. These potentials can
 be combined with models of three-nucleon force.  The state-of-the-art 3NF's are refined versions 
of the   Fujita-Miyazawa force ~\cite{Fuj57}, in which one of the nucleons is excited into
 intermediate $\Delta$ via 2$\pi$-exchange with both nucleons. In the general case  a pion 
 emitted by one nucleon interacts with a second nucleon and then is absorbed by a third nucleon.
 The modern version of the 2$\pi$-exchange Tucson-Melbourne (TM) 3NF model 
 \cite{Coo79, Coo81, Glo81}, called TM99 3NF,  is consistent with chiral symmetry~\cite{Coo01}. 
 It contains only one cut-off parameter, $\Lambda_{TM}$. 
 The value of $\Lambda_{TM}$ is adjusted  to reproduce the value of the ${^3}H$ 
binding energy~\cite{Nog97}. When the 3$N$ system  dynamics is studied with the AV18 NN potential also the Urbana~IX 3NF~\cite{Pud97} can be used. This force contains the 
two-pion exchange contribution due to intermediate $\Delta$ excitation  supplemented 
by a purely phenomenological repulsive short-range part. 

The other approach extends the nucleon-nucleon interaction picture to non-nucleonic 
degrees of freedom within the Coupled-Channels Potential (CCP). It is based on the 
realistic CD Bonn potential, but extended to include the $\Delta$-isobar as an active degree of freedom \cite{Del03, Del03a}.  In the energy range below the pion-production 
threshold, where the  $\Delta$-isobar excitation is virtual, it is assumed to be a stable 
baryon with real mass of 1232\,MeV. The CCP is based on the exchange of $\pi$, $\rho$, $\omega$,
 $\sigma$ mesons, with contribution of the transition between the NN and N$\Delta$ states, as well as 
 the exchange N$\Delta$-$\Delta$N potential. For the 3N system virtual excitation of 
 $\Delta$-isobar yields an effective 3NF. There is also a contribution to the transition amplitude of the so-called two-baryon dispersion in NN system. These two contributions usually compete, therefore the net  effects of including $\Delta$-isobar are suppressed as compared to the effects of the model 3NFs.\\
 \indent
The most extensive developments of nuclear potentials are nowadays carried out within the 
framework of the Chiral  Perturbation Theory  (ChPT). This effective field theory bases 
on the most general Lagrangian  for Goldstone bosons (pions) and matter fields (nucleons) 
 consistent with the broken chiral symmetry of the QCD \cite{Bed02, Epe09}. Resulting interaction consists of  long range and medium range pions exchanges, 
 and contact interaction, with the associated low energy constants. In the framework of ChPT the nuclear potential is obtained in a way of 
 a systematic expansion in terms of momentum variable: ($Q/\Lambda$)$^{\nu}$, 
 where $Q$ refers to a momentum of the nucleons,
 $\Lambda$ is connected to the chiral symmetry breaking scale and $\nu$ encounters expansion order. One of the most important features of the ChPT is the possibility to derive consistent many-body forces on the top of two-body ones. The first non-vanishing 3NF terms appear 
 in the next-to-next-to-leading  order (N2LO, 3rd order of chiral expansion, $\nu$=3). Recently, 
  the new version of chiral NN potential  has been developed, with  an improved semi-local regularization framework \cite{Epe15, Rei18}. In addition a new method of quantification of uncertainty due to the truncation  of the chiral expansion has been proposed \cite{Epe15a, Bin18}.  The possibility of estimating  the theoretical  uncertainties of the obtained predictions is an exceptional advantage as compared 
  to other approaches, see also Ref.~\cite{Ski18}.
It has been shown, that with regard to NN interactions it is necessary to perform calculations at 5th 
 order ($\nu$=5), {\it i.e.} N4LO (see the discussion in \cite{Mac16}). Thus, the 3NF at the same order is required for the consistency. So far the complete calculations for Nd system at N3LO are unavailable.    
  That is why  only approaches based on
   the realistic potentials are considered in this work.

\subsection{Relativistic effects}
 
Until recently, Faddeev calculations of observables in the deuteron breakup process  were carried out 
in a non-relativistic framework. The relativistic treatment of the breakup reaction in 3N system 
is quite a~new achievement  \cite{Ski06, Wit11}. 

From the theoretical point of view, the dynamical relativistic effects taken into account are the boost of  NN potential, relativistic deuteron wave function and  form of Lippmann-Schwinger equation and proper treatment of  Wigner rotations of spin states. Kinematical effects coming from relativistic phase-space factor are also included. The relativistic effects reveal at different parts
 of the breakup phase-space with various magnitude. 
 The calculations for the $^2H(n,nn)p$ breakup reaction showed that the relativistic effects tended to localize in phase-space
 regions characterized by small kinetic energy  of the undetected  proton  and simultaneously  the coplanarity of two neutrons  ($\varphi_{12}\approx180^{0}$)~\cite{Ski06}.
The relativity can increase or decrease,
 depending on the phase-space region, the nonrelativistic cross section and  magnitude of the effect
 increases with growing  neutron energy.  While at 65 MeV the influence of relativity effects is rather
 moderate ($\sim 20\%$) at 200 MeV they can change the nonrelativistic cross section even by 
 a factor of $\sim 2$.

\subsection{Coulomb interaction}

With the aim to incorporate Coulomb interaction in calculations for proton-deuteron collisions,  
the Coulomb potential is  screened and resulting scattering amplitudes are corrected 
 by renormalization technique to match  the unscreened limit. At first, 
 the Coulomb interaction  was applied to a purely nucleonic CDB
  potential and its coupled-channel extensions, CD Bonn+$\Delta$  \cite{Del05a,Del06}. 
 In the next step, the Coulomb force was  implemented in calculations   with the
  realistic AV18 NN potential combined with the Urbana IX three-nucleon
   force~\cite{Del09}. In this way the Coulomb and 3NF effects can be studied not only separately but also
  together what allows us to understand their interplay in the deuteron-proton data.

\section{\label{secIV}Results}

The differential cross section for a regular grid of polar and azimuthal angles with a constant step in arclength variable $S$ is obtained according to Eq.~(\ref{eq_break}). Polar angles of the two protons $\theta _1$ and $\theta _2$ are changed between 5$^\circ$ and 15$^\circ$ with the step size of 2$^\circ$ and their relative azimuthal angle  $\varphi _{12}$ is analszed in the~range from 20$^\circ$ to 180$^\circ$, with the step size of 20$^\circ$. In total, 189 configurations have been analysed. For each combination of the central values $\theta _1$, $\theta _2$, and $\varphi _{12}$ the experimental data are  integrated within the limits of $\pm$1$^\circ$ for the polar angles and  of $\pm$5$^\circ$ for relative azimuthal angle.  The bin size along the kinematic curve $S$ is either  8\,MeV or 24\,MeV, depending on the data rate in this region, in order to obtain statistical uncertainty per data point below 2\% in the maximum of the $S$ distribution.  

\begin{table}[h]
\caption{Definition of abbreviations applied for naming theoretical calculations. ``aver'' means averaging over angular ranges accepted in data analysis.}\label{theo}
\vspace{-2mm}
\begin{small}
\begin{center}
\begin{tabular}{|l|l|c|l|}
\hline
abbreviation  & description & aver & Ref. \\
\hline
\hline
 &  potentials:  &  & \cite{Mac01}\\
2N &  AV18, CD Bonn,  & Yes & \cite{Wir95}\\\
& Nijmegen I and II & & \cite{Sto94}\\
\hline
 &  potentials  (as above) &  & \cite{Coo79}\\
2N+TM99 & with TM99 3NF  & Yes & \cite{Coo81}\\
&   & &\cite{Glo81} \\
\hline
&  &  & \\
CDB &  CD Bonn potential & Yes & \cite{Mac01} \\
 &   & & \\
\hline
& coupled-channel potential&  & \\
CDB+$\Delta$ & CD Bonn+$\Delta$  & Yes & \cite{Del03, Del03a} \\
&  &  & \\
\hline
& coupled-channel potential &  & \\
CDB+$\Delta$+C &  CDB+$\Delta$   & Yes & \cite{Del05a, Del06}\\
 & with Coulomb force   & & \\
\hline
&CD Bonn potential &  & \\
CDBrel & relativistic  & No & \cite{Ski06}\\
 &  calculations  &  &\\
\hline
\end{tabular}
\end{center}
\end{small}
\end{table}

The data are compared with the theoretical calculations listed in the Table~\ref{theo}.
Prior to comparing with the data, a majority of the theoretical predictions has been averaged 
over the angular ranges accepted in the data analysis ($\Delta\theta_{1,2}=2^{\circ}$, 
$\Delta\varphi_{12}=10^{\circ}$) and projected onto relativistic kinematics, see 
Ref.~\cite{Kis13}. Relativistic 
calculations are the only exception: the calculations are performed for central values of 
the angular ranges alone. The theoretical calculations
using standard semi-phenomenological two-nucleon potentials, denoted in following NN,
provide very similar results and are treated as a~group: they are presented 
in figures as bands and, in 
calculations of $\chi^2$/d.o.f., an average value of all predictions is taken 
(corresponding to the middle of the band). Calculations using those 
potentials combined with the TM99 3NF (2N+TM99) are treated in an analogous 
way. The group of calculations, 2N, 2N+TM99, CDBrel, is performed with $np$ interaction in 1S0 wave, while the second group, CDB, CDB+$\Delta$ and CDB+$\Delta$+C, is performed using both $pp$ and $np$ interactions in all isospin triplet waves, including 1/2 and 3/2 total 3N isospin components.

Figs.~\ref{fig5t5}~-~\ref{fig15t15} present examples of the  differential cross section obtained for the chosen kinematic 
configurations  of the breakup reaction (at the beam energy of 170 MeV/nucleon).
 Each of figures shows
the set of experimental data compared to two different groups of theoretical calculations. In the top part the effects of 3NF (due to explicit treatment of $\Delta $ isobar) and influence of Coulomb interaction are presented. In the bottom part the effects of TM99 3NF  are shown. Fig.~\ref{fig5t5} presents configurations characterised with the lowest 
(among all analysed) proton polar angles ($\theta _1=\theta _2=5^\circ$), Fig.~\ref{fig15t15} - configurations with the largest proton polar angles ($\theta _1=\theta _2=15^\circ$), and Fig.~\ref{fig9t13} - sample configurations with asymmetric combination of proton polar angles ($\theta _1$=9$^\circ$, $\theta _2$=13$^\circ$). 
 In the figures, the error bars represent the statistical uncertainties, often smaller than the 
 data points. The systematic 
 uncertainties are represented by hatched bands in the lower part of each individual panel. Full set of data has been presented in Ref.~\cite{Klo17}.

\subsection{$\chi^2$ analysis}

Quantitative analysis of the description of the cross section data ($\sigma_{exp}$) 
provided by various calculations ($\sigma_{teor}$) is performed in 
terms of $\chi^2$-like variables. Due to dominating 
contribution  of systematic uncertainties, the following definition has been applied:

\begin{equation}
\chi^2 = \frac{1}{n_{d.o.f.}} \sum \frac{(\sigma_{teor}(\xi)-\sigma_{exp}(\xi))^2}{(\Delta\sigma_{st}(\xi)+\Delta\sigma_{sys}(\xi))^2},
 \label{chi2}
\end{equation}
where $\xi$ represents a set of kinematic  
variables $\xi$=($\theta_{1}$,$\theta_{2}$,$\varphi_{12}$,$S$), $\Delta\sigma_{st}(\xi)$ and  
$\Delta\sigma_{sys}(\xi)$ denote statistical and systematic uncertainties, respectively, 
summing goes over certain set of kinematic variables and $n_{d.o.f.}$ is a number of 
data points included in this sum. 
So defined quantity has no precise statistical meaning, however, it is still a measure of 
description provided by different models. When its value reaches 
roughly 2 or more, it can be treated as a signal of inconsistency 
between the model predictions  and the measured data. 
 
\begin{figure*}[h]
\includegraphics[width=12.9cm]{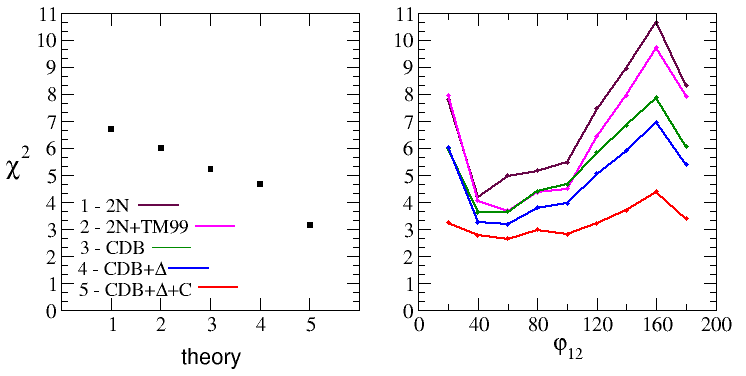}

\vspace{-0.5cm}

\caption{\label{globalchi2}
(Color online) Quality of description of the differential cross section for the breakup reaction at 170\,MeV/nucleon beam energy at forward angles. Left panel: The global $\chi^2/d.o.f.$ obtained as a result of 
comparing the cross section data 
with each of five types of theoretical calculations specified  in the legend. Right panel: The $\chi^2/d.o.f.$ calculated for set of data characterised with given relative azimuthal angle $\phi_{12}$}.%
\end{figure*}

The $\chi^2$ per degree of freedom defined above is calculated  globally, 
individually for the kinematic configurations  and, in addition, for the 
data sorted according to combination of polar angles $\theta_1, \theta_2$ and to relative 
azimuthal $\phi_{12}$ of the two protons.
Global analysis (see Fig.~\ref{globalchi2}, left panel) shows the importance of Coulomb
 interaction in the 
studied region of phase space.  Global analysis indicates also certain 
improvement of description due to including of 3NF, both for TM99 
force and in explicit $\Delta$ isobar approach.

The analysis performed in function of $\phi_{12}$ (see Fig.~\ref{globalchi2}, right panel) 
indicates clearly the region of dominance of the Coulomb effect. As expected, 
the region of the lowest  $\phi_{12}$, close to proton-proton FSI, is particularly
sensitive to Coulomb interaction, which lowers cross section by a large factor 
(see also configuration  $\theta _1$=5$^\circ$, $\theta _2$=5$^\circ$, $\varphi_{12}$=20$^\circ$ in 
Fig.~\ref{fig5t5}, top panel). The opposite 
influence of Coulomb interactions is present at the largest  $\phi_{12}$ (e.g. configuration  $\theta _1$=5$^\circ$, $\theta _2$=5$^\circ$, $\varphi_{12}$=180$^\circ$ in Fig.~\ref{fig5t5}), which is also 
visible as an increase of $\chi^2$ in Fig.~\ref{globalchi2}, right panel. The region 
of intermediate angles of about 60$^\circ$-80$^\circ$ is less sensitive to Coulomb repulsion between 
protons, though even there the effects are not negligible. In this region, the effect
of 3NF shows up   - not because of its particular strength, but since it is not covered that 
much by Coulomb effects.  The improvement is similar in case of 2N+TM99 and  CDBonn +$\Delta$ potentials.

\begin{center}
\begin{figure*}[h]
\includegraphics[width=14.7cm]{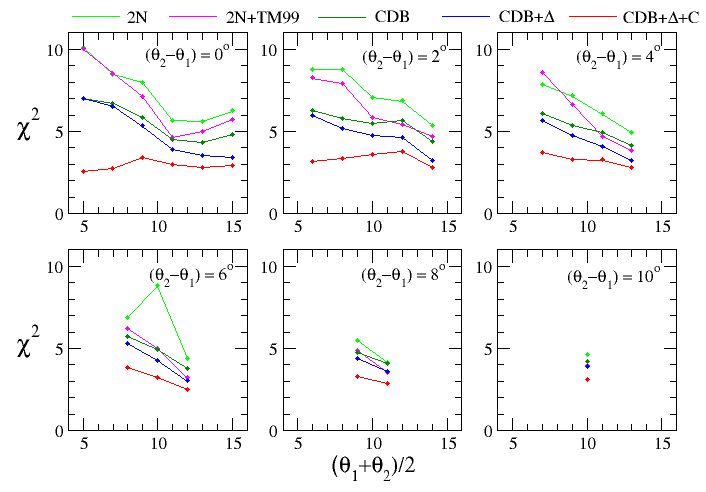}%

\vspace{-0.5cm}

\caption{\label{chi2_theta}
(Color online) Similar to analysis shown in Fig.~\ref{globalchi2}, right panel, but with 
$\chi^2$ per degree of 
freedom  calculated for each set of data characterised with the given combination of proton polar 
angles.  The results are ordered according to difference of polar angles 
$\theta_{12}=\theta_1-\theta_2$; in each panel results for one value of $\theta_{12}$ are 
shown. 
}%
\end{figure*}
\end{center}

\begin{center}
\begin{figure*}[h]
\subfigure{
\includegraphics[width=14.2cm]{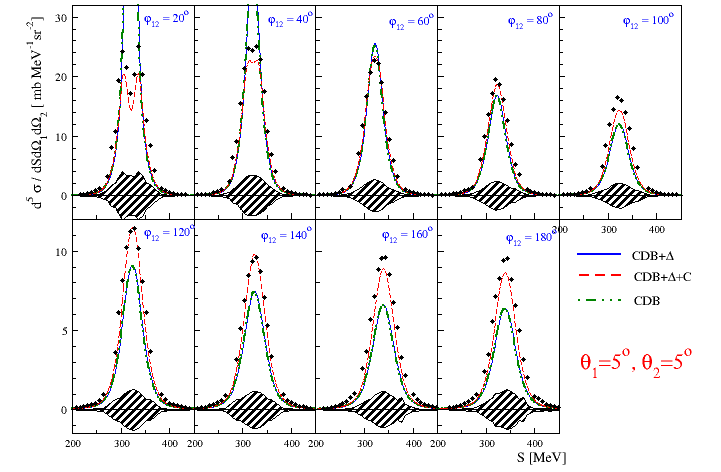}
}
\subfigure{
\includegraphics[width=14.2cm]{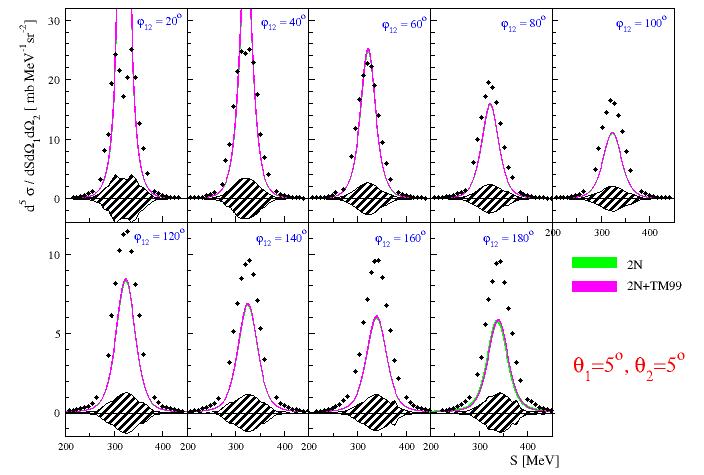}
}\\

\caption{(Color online) Differential cross section of $^{1}$H$(d,pp)n$ breakup reaction at beam energy of 170\,MeV/nucleon shown in function of $S$ variable. The presented data belong to 9 kinematic configurations characterised with the same combinations of proton polar angles ($\theta _1$=5$^\circ$, $\theta _2$=5$^\circ$) and various relative azimuthal angles of the two protons, indicated in the individual panels. Statistical errors are smaller than the point size. Systematic uncertainties are represented by hatched band. {\bf Top part:} data compared to predictions obtained for the calculations within the coupled-channel approach with the CD Bonn potential (CDB, dotted-dashed green line), with the CDBonn +$\Delta$ potential  without (CDB+$\Delta$, blue solid line) and with  the Coulomb force included (CDB+$\Delta$+C, red dashed line);  {\bf Bottom part:} the same data confronted with the the predictions based on NN potentials: 2N (AV18, CD Bonn, Nijm I and II) (green band) and NN combined with the TM99 3NF (magenta band).
\label{fig5t5}}
\end{figure*}
\end{center}

\begin{center}
\begin{figure*}[h]
\subfigure{
\includegraphics[width=14.8cm]{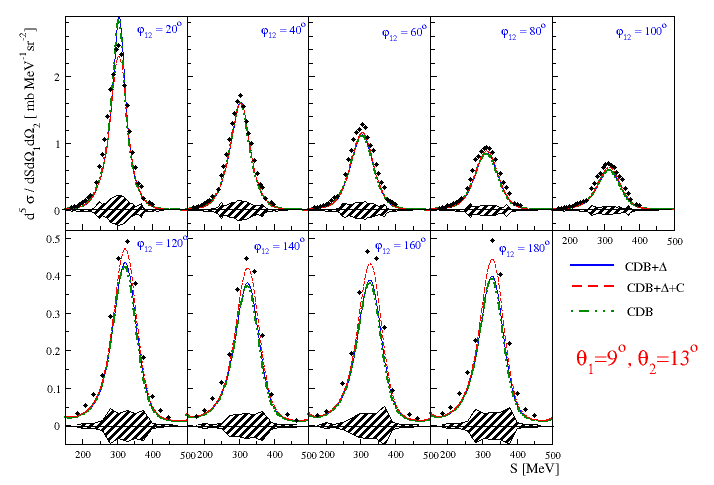}
}
\subfigure{
\includegraphics[width=14.8cm]{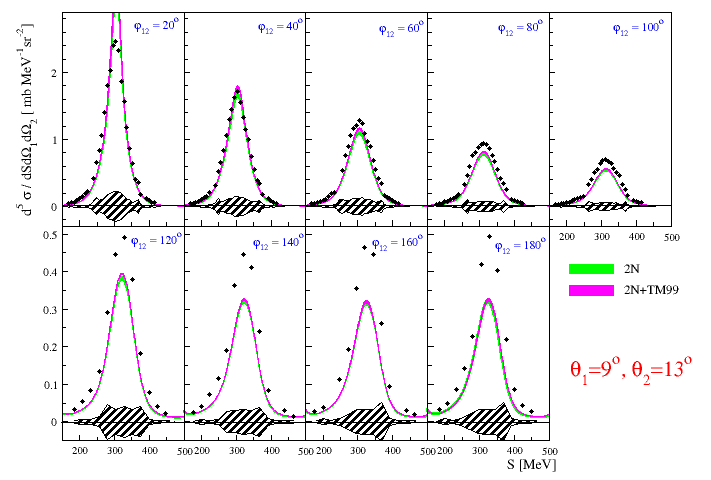}
}\\
\caption{(Color online) The same as Fig.~\ref{fig5t5} but for kinematic configurations with the proton polar angles  $\theta _1$=9$^\circ$, $\theta _2$=13$^\circ$.
\label{fig9t13}}
\end{figure*}
\end{center}

\begin{center}
\begin{figure*}[h]
\subfigure{
\includegraphics[width=14.8cm]{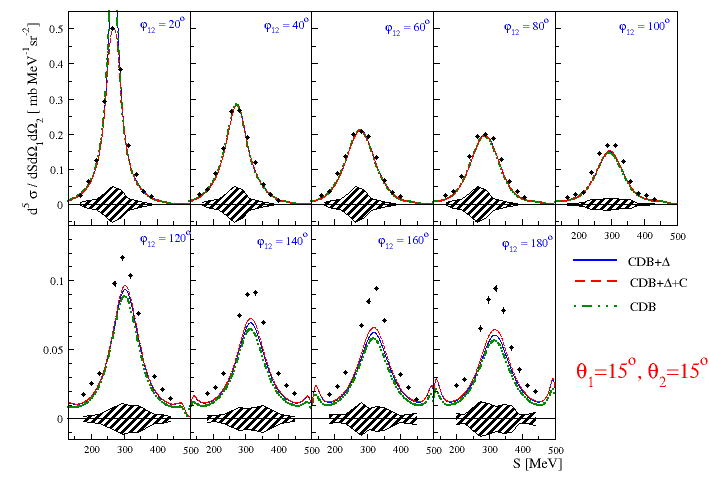}
}
\subfigure{
\includegraphics[width=14.8cm]{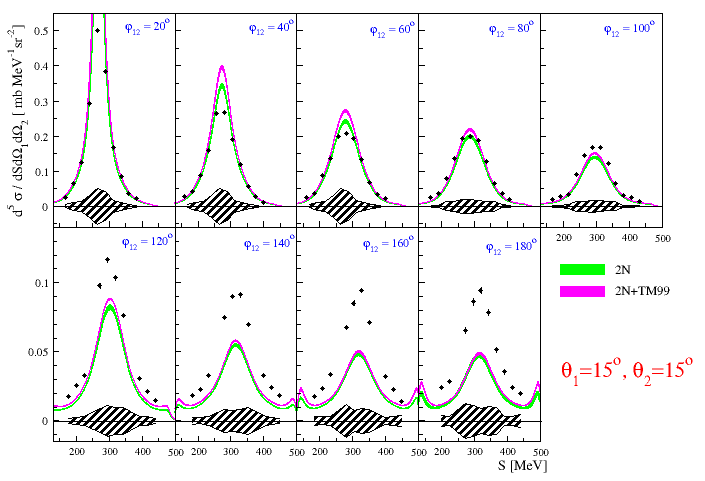}
}\\
\caption{(Color online) The same as Fig.~\ref{fig5t5} but for kinematic configurations with the proton polar angles  $\theta _1$=15$^\circ$, $\theta _2$=15$^\circ$.
\label{fig15t15}}
\end{figure*}
\end{center}

The analysis of data sorted in  function of  combination of polar angles $\theta_1, \theta_2$ 
(Fig.~\ref{chi2_theta}) 
provides another examples of the Coulomb force dominance in the  FSI region, characterised by  the lowest difference of polar angles. 

Dominant influence of Coulomb interaction at forward proton
emission angles (in laboratory system of the $^{1}$H$(d,pp)n$ reaction) is in agreement with 
studies at other beam energies, see for example \cite{Cie15}.

\subsection{Relativistic effects}

\begin{center}
\begin{figure*}[h]

\includegraphics[width=13.7cm]{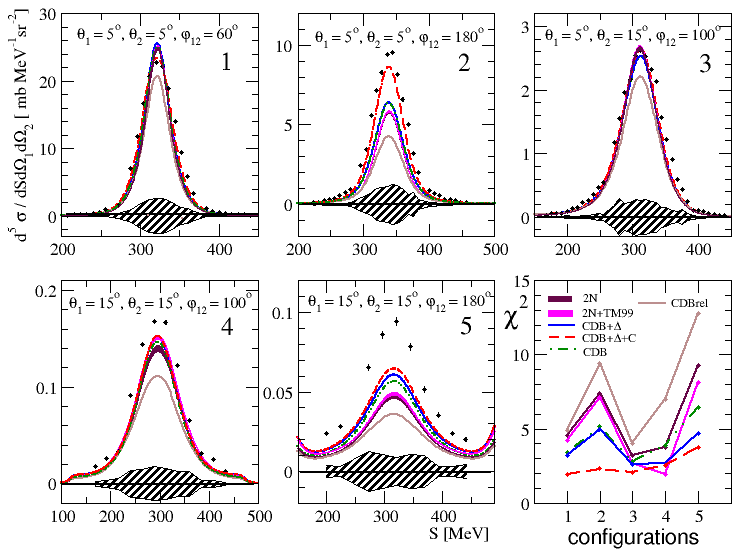}

\caption{(Color online) Similar to Fig.~\ref{fig5t5} but for the set of configurations for which relativistic 
NN calculations has been performed (shown as brown solid line).  The bottom right panel shows 
the $\chi^2/d.o.f.$ analysis performed for each configuration shown in this figure, numbered  as 
in the panels. The brown points (connected by solid brown line) represent results of $\chi^2/d.o.f.$ analysis with 
relativistic calculations.}
\label{kon2rel}      
\end{figure*}
\end{center}

Fig.~\ref{kon2rel} presents the set of configurations for which relativistic 
NN calculations have been performed.
The result presented in the top left panel indicates  an interplay of 3N interactions,
 Coulomb force 
and relativistic effects. For the configuration shown in the bottom panel all the 
calculations 
underestimate the experimental data. The discrepancy is even increased by relativistic 
calculations, which is also reflected in the $\chi^2$ analysis.

\section{\label{secV}Summary}

The differential cross section of the $^{1}$H$(d,pp)n$ reaction has been determined for the  
configurations characterised with forward proton emission angles in the laboratory system. 
In spite of 
the relatively high beam energy of 170 MeV/nucleon, the Coulomb interaction plays 
a dominant role in this region. The predicted 3NF effects are small or very moderate, nevertheless 
the description of the experimental data is improved by 
including the 3NF into calculations. It is observed in  both approaches applied to 
modelling the 3N force and the improvement is  seen in the region where the net Coulomb
 effects are moderate. 
None of the existing calculations, even the one including both 
the Coulomb interaction and three-
nucleon force, provides satisfactory description of the whole data set. 
The problem is observed at the largest studied polar angles of two protons:
 $\theta_{1},\theta_{2} \geq 13^{\circ}$ combined with their large relative azimuthal 
angle $\varphi_{12}>120^{\circ}$, where all the predictions underestimate the measured cross 
section. This effect can be associated with problem in describing the elastic scattering 
cross section at its minimum. On the other hand, the full relativistic treatment of the process is still 
missing.  The relativistic calculations based on pure NN interaction show the effect 
opposite to the one needed for the correct data description, 
but it will be interesting to see contributions of 3NF included in relativistic calculations. The data set
collected in the experiment under discussion contains  also the strongly asymmetric configurations:  
coincidences with one proton
 registered in  FD and the other one - in Central
 Detector (FD-CD), which correspond to the angular range 
 $5^\circ < \theta_{1} < 15^\circ$, $20^\circ < \theta_{2} < 90^\circ$. They will be used in forthcoming analysis to further 
 explore the observed situation, along with the data collected  at lower 
(170\,MeV/nucleon) and higher (190 and 200\,MeV/nucleon) deuteron beam energies. 

\begin{acknowledgments}
This work was partially supported by Polish National Science Center under Grants No. 2012/05/E/ST2/02313, No. 2012/05/B/ST2/02556 and No. 2016/22/M/ST2/00173. We thank the COSY crew for their work and the excellent conditions during the beam time.

\end{acknowledgments}

\end{document}